\documentclass[aps,prl,reprint,showpacs,showkeys,superscriptaddress,preprintnumbers,nofootinbib]{revtex4-1} 
\usepackage{amsmath,amssymb,bm,mathrsfs}
\usepackage{hyperref}
\usepackage{url}
\usepackage{breakurl}
\usepackage{graphicx}
\usepackage{subcaption}

\def\simgt{\mathrel{\lower2.5pt\vbox{\lineskip=0pt\baselineskip=0pt
           \hbox{$>$}\hbox{$\sim$}}}}
\def\simlt{\mathrel{\lower2.5pt\vbox{\lineskip=0pt\baselineskip=0pt
           \hbox{$<$}\hbox{$\sim$}}}}

\begin{document}

\title{{Higgsino Dark Matter or Not:} \\ 
{\normalsize Role of Disappearing Track Searches at the LHC and Future
Colliders}} 

\author{Hajime Fukuda}
\email{hajime.fukuda@ipmu.jp}
\affiliation{
Kavli Institute for the Physics and Mathematics of the Universe (WPI),
The University of Tokyo Institutes for Advanced Study, The University of
Tokyo, Kashiwa 277-8583, Japan
}

\author{Natsumi Nagata}
\email{natsumi@hep-th.phys.s.u-tokyo.ac.jp}
\affiliation{Department of Physics, University of Tokyo, 
Tokyo 113-0033,
Japan}

\author{Hidetoshi Otono}
\email{Hidetoshi.Otono@cern.ch}
\affiliation{
Research Center for Advanced Particle Physics, Kyushu University,
Fukuoka 819-0395, Japan 
  }

\author{Satoshi Shirai}
\email{satoshi.shirai@ipmu.jp}
\affiliation{
Kavli Institute for the Physics and Mathematics of the Universe (WPI),
The University of Tokyo Institutes for Advanced Study, The University of
Tokyo, Kashiwa 277-8583, Japan
}

\begin{abstract}

 Higgsino in supersymmetric standard models is known to be a promising
 candidate for dark matter in the Universe. Its phenomenological
 property is strongly affected by the gaugino fraction in the
 Higgsino-like state. If this is sizable, in other words, if gaugino
 masses are less than ${\cal O}(10)$~TeV, we may probe the Higgsino dark
 matter in future non-accelerator experiments such as dark matter direct
 searches and measurements of electric dipole moments. On the other
 hand, if gauginos are much heavier, then it is hard to search for
 Higgsino in these experiments. In this case, due to a lack of gaugino
 components, the mass difference between the neutral and charged
 Higgsinos is uniquely determined by electroweak interactions to be
 around $350$~MeV, which makes the heavier charged state rather
 long-lived, with a decay length of about $1$~cm. In this
 letter, we argue that a charged particle with a flight length of ${\cal
 O}(1)$~cm can be probed in disappearing-track searches if
 we require only two hits in the pixel detector. Even in this case, we
 can reduce background events with the help of the displaced-vertex
 reconstruction technique.  We study the prospects of this search
 strategy at the LHC and future colliders for the Higgsino dark
 matter scenario. It is found that an almost pure Higgsino is indeed
 within the reach of the future $33$~TeV collider experiments. We then discuss that
 the interplay among collider and non-accelerator experiments plays a
 crucial role in testing the Higgsino dark matter scenarios. Our
 strategy for disappearing-track searches can also enlarge the discovery
 potential of pure wino dark matter as well as other electroweak-charged
 dark matter candidates.

\end{abstract}

\maketitle
\preprint{IPMU 17-0050}
\preprint{UT-17-11}
\preprint{KYUSHU-RCAPP-2017-03}

%%%%%%%%%%%%%%%%%%%%%%%
\section{Introduction}
%%%%%%%%%%%%%%%%%%%%%%%

The strongest motivation for new physics beyond the Standard Model (SM)
is dark matter (DM) in the Universe. Among a variety of DM candidates,
weakly-interacting massive particles (WIMPs) offer the most attractive
paradigm to explain the observed DM density: $\Omega_{\rm DM} h^2 \simeq
0.12$~\cite{Ade:2015xua}.

Higgsino in a supersymmetric (SUSY) extension of the SM, which is an
SU(2)$_L$ doublet fermion with hypercharge $1/2$, is a prime example of
WIMP DM. The neutral component of Higgsino can mix with the neutral
electroweak-inos, bino and wino, and if the dominant portion of the
lightest state is Higgsino, then it is called a Higgsino-like state. The
amount of the gaugino fractions in a Higgsino-like state strongly
affects its phenomenological property. If the masses of SUSY particles
other than Higgsino, especially those of the electroweak-inos, are in
the multi-TeV region, then the gaugino fraction is quite suppressed and
thus the lightest SUSY particle (LSP) can be regarded as an almost pure
Higgsino. Such a ``split'' mass spectrum has various phenomenological
and theoretical advantages~\cite{Wells:2003tf, *Wells:2004di,
ArkaniHamed:2004fb, *Giudice:2004tc, *ArkaniHamed:2004yi,
*ArkaniHamed:2005yv}, and thus attracts wide
attention \cite{Hall:2011jd,*Hall:2012zp,*Nomura:2014asa, Ibe:2011aa,
*Ibe:2012hu, Arvanitaki:2012ps, ArkaniHamed:2012gw, Evans:2013lpa,
*Evans:2013dza, *Evans:2014pxa} especially after the discovery of the 
125~GeV Higgs boson \cite{Aad:2012tfa, *Chatrchyan:2012ufa, *Aad:2015zhl}.
The gaugino fraction may be small, but non-accelerator experiments such as DM direct searches and measurements of the electric dipole moments (EDMs) can probe it. As
we discuss below, even the gauginos heavier than $\mathcal O(10)\,\text{TeV}$ can be probed in future
experiments \cite{Nagata:2014wma}.

Another motivation for SU(2)$_L$ doublet DM is provided by
a bottom-up approach. Classifying possible DM candidates
in terms of their quantum numbers, we find that an SU(2)$_L$ doublet
with hypercharge $1/2$ or triplet with hypercharge zero has minimal
non-zero electric charge to accommodate electrically neutral DM.
Hence, SU(2)$_L$ doublet DM is a promising target even
though we do not consider SUSY as new physics beyond the SM.

However, SU(2)$_L$ doublet fermion DM as it is, which
corresponds to completely pure Higgsino DM, has already been
excluded by DM direct detection experiments. In this case, the DM
is a Dirac fermion, and thus has vector interactions with quarks
via the $Z$-boson exchange, which results in a too large 
DM-nucleon spin-independent (SI) scattering cross section \cite{Akerib:2016vxi, *Tan:2016zwf}. 
To avoid this constraint, we need some effects induced at higher energies that 
generate a mass split between the Dirac DM components, $\Delta
m_0$, to divide it into two Majorana fermions $\widetilde
\chi^0_{1,2}$, which do not have vector interactions. To assure this, we
need $\Delta m_0 \gtrsim \mathcal O(100)\,\text{keV}$
\cite{Nagata:2014wma}; otherwise, the inelastic scattering process
$\widetilde\chi^0_{1}N \to \widetilde\chi^0_{2} N$ ($N$ denotes a
nucleon) instead gives a too large scattering cross
section. Generically, the interactions that yield $\Delta
m_0$ should be induced at a scale $\Lambda \lesssim 10^9$~GeV to give
$\Delta m_0 \gtrsim \mathcal O(100)\,\text{keV}$ \cite{Nagata:2014wma,
Nagata:2014aoa}. Such effects are actually similar to the contribution of
gaugino fractions to the property of the Higgsino DM. Thus, this
$\text{SU}(2)_L$ DM can also be probed in future non-accelerator
experiments if the scale $\Lambda$ is relatively low. Notice that bounds
from these experiments give a lower limit on $\Lambda$, while the
inelastic-scattering bound gives an upper limit.

We however find that there is a range of $\Lambda$
which is high enough for the SU(2)$_L$ doublet fermion DM to be beyond
the reach of these experiments but low enough to give $\Delta m_0
\gtrsim \mathcal O(100)\,\text{keV}$ \cite{Nagata:2014wma, Nagata:2014aoa}---correspondingly, we can find appropriately large
gaugino masses with which the same situation is realized for
Higgsino DM. This situation is completely viable, as the
thermal relic abundance of pure Higgsino DM agrees with $\Omega_{\text{DM}} h^2 \simeq 0.12$ if the Higgsino DM mass $m_{\text{DM}}$ is $\sim 1.1$~TeV
\cite{Cirelli:2007xd}. A
lighter mass region is also promising since there may be non-thermal
contribution from high-energy physics, such as late-time decay of
gravitinos. Therefore, it is quite important to consider an alternative
strategy to probe such almost pure Higgsino DM.

It has been known that almost pure Higgsino is also difficult to probe
at hadron colliders \cite{Han:2014kaa, Ismail:2016zby}. In this case,
not only the neutral states 
$\widetilde \chi^0_{1,2}$, but also the charged states
$\widetilde\chi^\pm$ are almost degenerate in mass at tree level. The
charged states acquire a slightly heavier mass via the electroweak loop
corrections, but still the mass splitting is as small as $\sim
350$~MeV. Due to this small mass splitting, the decay products of
$\widetilde\chi^\pm$ have soft momenta, which makes it hard to
detect them. For this reason, the current LHC searches for Higgsinos
rely only on mono-jet or mono-$X$ signals, which give poor constraints
on Higgsinos due to large SM background (BG). Even a 100~TeV collider
can test a small portion of parameter space in the Higgsino DM
scenario \cite{Low:2014cba, *Gori:2014oua, *Arkani-Hamed:2015vfh,
*Golling:2016gvc}.

However, this small mass splitting may offer a different strategy to
probe Higgsinos at colliders. Because of the small mass splitting, the
charged Higgsino has rather long lifetime---as it turns out, a mass
difference of $\sim 350$~MeV results in a decay length of $\sim 1$~cm. It is
therefore interesting to ask whether charged tracks with an ${\cal
O}(1)$~cm length can be detected at collider experiments. In this
letter, we address this question by taking into account potential
improvements in disappearing track searches with fully
optimized inner detectors at the LHC and a future 33~TeV collider. Such
an improvement may also increase the discovery potential of pure wino
DM; thus, we also consider the wino DM case for
comparison.

%%%%%%%%%%%%%%%%%%%%%%%%%%%%%%%%%%%%%%%%%
\section{Higgsino Phenomenology}
%%%%%%%%%%%%%%%%%%%%%%%%%%%%%%%%%%%%%%%%%

To begin with, we discuss the mass spectrum of Higgsino components,
paying particular attention to the gaugino mass effects on it. 
The masses of these components split after the electroweak symmetry is
broken. These mass splittings are induced by the tree-level mixing with the
electroweak-inos, as well as the electroweak radiative corrections. 
We here assume that the masses of the electroweak-inos are much larger
than the TeV scale. 
At one-loop level, the electroweak-loop contribution to the
charged-neutral mass splitting is given by 
\begin{equation}
\Delta m_{\text{rad}} \simeq \frac 12 \alpha_2 m_Z\sin^2\theta_w
\left(1 -\frac{3m_Z}{2\pi m_{\widetilde\chi^\pm}}\right) ~,
\end{equation}
where $\alpha_2\equiv g_2^2/4\pi$ with $g_2$ the SU(2)$_L$ gauge
coupling constant, $m_Z$ is the $Z$-boson mass,
$\theta_w$ is the Weinberg angle, and $m_{\widetilde{\chi}^\pm}$ is the
charged Higgsino mass. We find $\alpha_2 m_Z \sin^2 \theta_w/2
\simeq 355$~MeV. After all, the mass difference between the neutral
components, $\Delta m_0$, and that between the charged component and the
Higgsino LSP, $\Delta m_+$, are respectively given by
\begin{eqnarray}
\label{eq:massdiff0}
\Delta m_0 &\simeq& \frac{v^2}{4|\mu|}|X| \Delta_X ~,\\ 
\label{eq:massdiffpm}
\Delta m_{+} &=& \Delta m_{\text{tree}} + \Delta m_\text{rad} ~,
\end{eqnarray}
with
\begin{equation}
 \Delta m_{\text{tree}} \simeq \frac{v^2}{8|\mu|} 
\Bigl[
|X| \Delta_X + \sin2\beta\, \text{Re}(Y)
\Bigr]
\, ,
\end{equation}
where $v\simeq 246$~GeV is the Higgs VEV, $\mu$ is the Higgsino mass,
$\tan\beta \equiv \langle H_u \rangle/\langle H_d \rangle$ with $H_u$
and $H_d$ being the up- and down-type Higgs fields, respectively,
\begin{equation}
 X\equiv \mu^*\left(\frac{g_1^2}{M_1} + \frac{g_2^2}{M_2}\right)~,
~~~
 Y\equiv \mu^*\left(\frac{g_1^2}{M_1} - \frac{g_2^2}{M_2}\right)~,
\end{equation}
$\Delta_X \equiv \sqrt{1-\sin^2\theta_X \sin^2 2\beta}$,
$\theta_{X} \equiv \text{arg}(X)$,
$M_1$ ($M_2$) is the bino (wino) mass, and $g_1$ is the $\text{U}(1)_Y$
gauge coupling. As we discussed above, $\Delta m_0
\gtrsim \mathcal O(100)\,\text{keV}$ is required to suppress the SI
inelastic DM-nucleon scattering via the $Z$-boson exchange.

Even in this case, the Higgsino-nucleon SI scattering occurs via
the Higgs-boson exchange process induced by the electroweak-ino
mixing. The resultant SI scattering cross section with a proton, for
instance, is evaluated as  
\begin{equation}
 \sigma_{\text{SI}}^{(p)}
\simeq \frac{4m_{\text{red}}^2 m_p^2
}{\pi v^4 m_h^4} 
F^2_{\text{mat}} \Delta m_{\text{tree}}^2
F(X,Y)^2
~,
\label{eq:sigsi}
\end{equation}
with 
\begin{equation}
 F(X,Y) \equiv \frac{|X| \Delta_X + \sin2\beta\, \text{Re}(X)}
{|X| \Delta_X + \sin2\beta\, \text{Re}(Y)} ~,
\end{equation}
where $m_p$ is the proton mass, $m_h \simeq 125$~GeV is the Higgs mass,
$m_{\text{red}} \equiv m_p m_{\text{DM}}/(m_p + m_{\text{DM}})$,
$F_{\text{mat}} \equiv \sum_{q=u,d,s} f_{T_q} +2f_{T_G}/9$, $f_{T_G}
\equiv 1- \sum_{q=u,d,s} f_{T_q}$, and the mass fractions $f_{T_q}$ are
computed with a QCD lattice simulation as $f_{T_u} = 0.0149$, $f_{T_d} =
0.0234$, and $f_{T_s} = 0.0440$ \cite{Abdel-Rehim:2016won}. In the
derivation of Eq.~\eqref{eq:sigsi}, we have dropped the electroweak-loop
contribution to the SI scattering cross section, which is found to be
negligible for almost pure Higgsino DM \cite{Hisano:2015rsa}. Using this
expression, we find a relation between the SI scattering cross section
and the tree-level charged-neutral mass splitting $\Delta
m_{\text{tree}}$: 
\begin{equation}
 \Delta m_{\text{tree}} \simeq 
\frac{170~\text{MeV}}{F(X,Y)} 
\left(
\frac{m_p}{m_{\text{red}}}
\right)
\biggl(
\frac{\sigma_{\text{SI}}^{(p)}}{10^{-48}~\text{cm}^2}
\biggr)^{\frac{1}{2}}~.
\end{equation}
This equation shows that for a value of $\sigma_{\text{SI}}^{(p)}$ that
exceeds the neutrino floor \cite{Billard:2013qya}, which lies around
$10^{-48}~\text{cm}^2$ for a DM mass of ${\cal O}(100)$ GeV, the
charged-neutral mass splitting is expected to be larger than a few
hundred~MeV.

In general, the electroweak-ino contribution yields CP violation, which
then induces EDMs of the SM particles. Currently, the electron EDM,
$d_e$, is most stringently restricted: $|d_e| < 8.7 \times 10^{-29}~e\text{cm}$
\cite{Baron:2013eja}. The Higgsino contribution to the
electron EDM is induced by the two-loop Barr--Zee diagrams
\cite{Barr:1990vd, Chang:2005ac, *Deshpande:2005gi, *Giudice:2005rz} and
given by the sum $d_e = d_{e}^{h\gamma} + d_e^{hZ} +
d_e^{WW}$, where $d_{e}^{h\gamma}$, $d_e^{hZ}$, and $d_e^{WW}$ are the
Higgs-$\gamma$, Higgs-$Z$, and $WW$ loop contributions, respectively. 
We find \cite{Nagata:2014wma}
\begin{align}
 d_e^{h\gamma} &\simeq \frac{e g_2^2 m_e \sin 2\beta}{(4\pi)^4
v^2|\mu|}
\biggl(
\frac{8\sin^2\theta_w f_0^h \text{Im}(X-Y)}
{|X| \Delta_X + \sin2\beta\, \text{Re}(Y)}
\biggr) 
\Delta m_{\text{tree}}
~,\nonumber \\
d_e^{WW} &\simeq
\frac{e g_2^2 m_e \sin 2\beta}{(4\pi)^4
v^2|\mu|}
\biggl(
\frac{f_0^W \text{Im}(Y)}
{|X| \Delta_X + \sin2\beta\, \text{Re}(Y)}
\biggr)
\Delta m_{\text{tree}}
~,
\label{eq:edm}
\end{align} 
with $f_0^{h,W} = f_0(|\mu|^2/m_{h,W}^2)$ and
\begin{equation}
 f_0(r) = r\int_0^1  \frac{dx}{r - x(1 - x)}
  \ln\left(\frac{r}{x(1-x)}\right)
\,,
\end{equation}
where $m_e$ is the electron mass, $m_W$ is the $W$-boson mass,  
and $e$ is the positron charge. For the
electron EDM, $d_e^{hZ}$ is always subdominant due to an accidentally
small numerical factor \cite{Nagata:2014wma}. Using this expression, we
can again relate $d_e$ with the mass splitting $\Delta m_{\text{tree}}$
as 
\begin{align}
 d_e \simeq
3\times 10^{-31}\cdot
 \sin(2\beta) 
\biggl(\frac{1~\text{TeV}}{|\mu|}\biggr)
\biggl(\frac{\Delta m_{\text{tree}}}{100~\text{MeV}}\biggr)
F_{\text{ph}}~e\text{cm}~,
\end{align}
where $F_{\text{ph}}$ is the sum of the factors in the parentheses in
Eq.~\eqref{eq:edm}. By noting that future experiments may probe $d_e
\sim 10^{-31}~e\text{cm}$ \cite{Vutha:2009ux, Hudson:2011zz}, we 
find that a Higgsino with $\Delta m_{\text{tree}} \gtrsim 100$~MeV can
be tested in EDM experiments.

As we have seen above, a Higgsino with $\Delta m_{\text{tree}} > {\cal
O}(100)$~MeV can be tested in future non-accelerator
experiments. On the other hand, an almost pure Higgsino with
$\Delta m_{\text{tree}}$ (much) smaller than $\sim 100$~MeV is beyond
the reach of these experiments. It is also challenging to probe such
Higgsinos in DM indirect searches, or mono-jet/$X$ searches at the LHC. 
However, as we mentioned above, in this case we may search
for Higgsinos using disappearing tracks at the LHC, since 
$\widetilde{\chi}^\pm$ becomes rather long-lived due to the
small mass difference $\Delta m_+$. The decay length of
$\widetilde{\chi}^\pm$ is approximately \cite{Chen:1995yu, *Thomas:1998wy} 
\begin{eqnarray}
\label{eq:ctau}
c\tau \simeq 0.7~\text{cm}\times \left[\left(\frac{\Delta
					m_+}{340\,\text{MeV}}\right)^3\sqrt{1
- \frac{m_\pi^2}{\Delta m_+^2}}\right]^{-1}, 
\end{eqnarray}
where $m_\pi$ is the charged pion mass. Hence, we expect that a pure charged Higgsino leaves an ${\cal
O}(1)$~cm track in the detector. We will see below that it is indeed
possible to detect an ${\cal O}(1)$~cm disappearing track at colliders.

%%%%%%%%%%%%%%%%%%%%%%%%%%%%%%%%%%%%%%
\section{Disappearing track searches}
%%%%%%%%%%%%%%%%%%%%%%%%%%%%%%%%%%%%%%

Disappearing track searches highly rely on the performance of silicon
pixel detectors. The CMS detector has three barrel layers of pixel
detectors in a magnetic field of 3.8~T at radii of 4.4~cm, 7.3~cm, and
10.2~cm with a pixel size of $100\times 150~\mu\text{m}^2$
\cite{Chatrchyan:2008aa}\,\footnote{Recently, the CMS collaboration has replaced the pixel detectors with four layers at radii 2.9~cm, 6.8~cm, 10.9~cm and 16.0~cm \cite{Tavolaro:2016hfj}.}. The ATLAS Pixel detector is located in a
magnetic field of 2~T and consists of four barrel layers at radii of
3.3~cm, 5.05~cm, 8.85~cm, and 12.25~cm \cite{PERF-2007-01}. The
innermost layer in the ATLAS Pixel detector system called Insertable
B-Layer (IBL) \cite{Capeans:1291633}, which was installed before the LHC
Run 2 started, has a pixel size of $50\times 250~\mu\text{m}^2$, while
the other layers have a pixel size of $50\times 400~\mu\text{m}^2$. For
concreteness, we mainly consider the ATLAS setup in this section.

The current disappearing track searches by the ATLAS
collaboration,\footnote{The CMS Run-1 study can be found in
Ref.~\cite{CMS:2014gxa}.}
whose main target is a long-lived charged wino, require four hits in
the silicon detectors. Thus, at the LHC Run 1, a target charged particle
needs to fly at least 29.9~cm, which corresponds to the location of the
innermost layer of the silicon strip detector called SemiConductor
Track (SCT), to pass this selection. This Run-1 analysis excludes a pure
charged wino with a mass of less than $270$~GeV \cite{Aad:2013yna}. Thanks to
the IBL, the required minimum flight length of target particles was
shorten (12.25~cm) at the Run 2, giving much better sensitivity to a
pure charged wino, which has a decay length of $\sim 6$~cm. The current
mass limit on a pure charged wino is $430$~GeV
\cite{moriond17, *moriond17conf}.

\begin{figure}
{\includegraphics[width=0.4\textwidth]{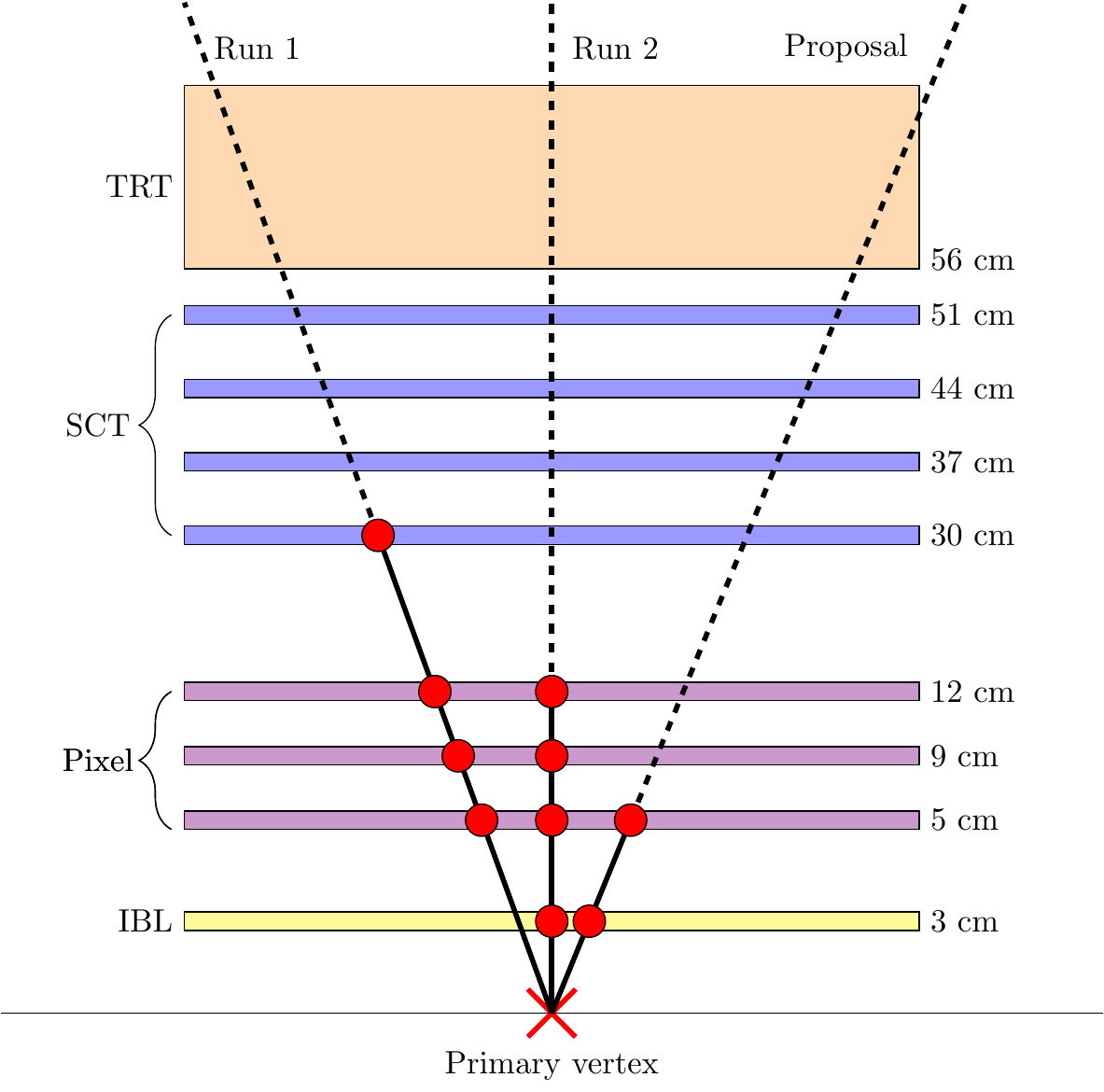}} 
\caption{
Required number of hits in the ATLAS inner tracker for the
analyses of Run-1\&2 and ours. 
}
\label{fig:track}
\end{figure}

A similar strategy may be adopted for charged Higgsino searches. As
shown in Eq.~\eqref{eq:ctau}, an almost pure charged Higgsino has a
decay length of $\sim 1$~cm, which suggests us to focus on shorter
tracks. To optimize the search strategy for this case, we here
require only two hits in the Pixel detector. In addition to these two
hits, a reconstructed primary vertex may be utilized to determine the
momentum of a disappearing track. Note that the resolution of the
position of a primary vertex is ${\cal O}(10)~\mu\text{m}$
\cite{ATL-PHYS-PUB-2015-026, *moriond17DV}, which is the same order as that of the
Pixel detector. Figure~\ref{fig:track} illustrates the configuration of
the ATLAS inner tracker, where required numbers of
hits in the detector are shown by red blobs for the analyses of the
ATLAS Run-1\&2 and ours. As shown in this figure, our search
strategy allows the ATLAS detector to probe a disappearing track
with a length of $\gtrsim 5$~cm.

According to the ATLAS Run-2 analysis \cite{moriond17, *moriond17conf}, there are three
classes of BG events in this search: hadrons that scatter
with the detector, leptons whose flight direction is bent with
Bremsstrahlung, and fake tracks due to mis-identification of hit
points. The total number of remaining BG events after the selection
cut is expected to be $11.8 \pm 3.1$ in the electroweak production
channel in the Run-2 study \cite{moriond17, *moriond17conf}. 
 
As we have reduced the number of hits in the silicon
detector, the resolution of the transverse momentum of a disappearing
track ($P^{\text{dis}}_{\text{T}}$) may be worse than the current
one, $\Delta (1/P_{\text{T}}^{\text{dis}})\sim 10~{\rm TeV}^{-1}$.
Since $\Delta (1/P_{\text{T}}^{\text{dis}}) \propto
L^{-2}$ with $L$ the track length, we expect $\Delta
(1/P_{\text{T}}^{\text{dis}})\sim 30~{\rm TeV}^{-1}$ for our two-hit
method. This means that if a disappearing track has
$P^{\text{dis}}_{\text{T}} \gtrsim 30$~GeV, its momentum can no longer
be measured accurately. This poor momentum resolution may increase the
number of BG events significantly. 

Meanwhile, these BG
events may be reduced with the help of reconstruction of displaced
tracks. This technique itself has already been established in the
long-lived particle searches with displaced vertices (DVs)
\cite{Aad:2011zb, *Aad:2012zx, *Aad:2015rba} with the current
reconstruction efficiency of displaced tracks from a radius of 5~cm
being about 80\% \cite{ATLAS:trk}. 
As the hadronic scattering and lepton Bremsstrahlung BG events
generally yield a kink signature, these events may be removed
efficiently if a DV having a disappearing
track and a displaced track is reconstructed. Notice that this requirement scarcely
reduces the signal events, since the emitted pion in the charged
Higgsino decay is too soft to be reconstructed as a kink signature. 
The fake-track events may also be resolved with the DV reconstruction,
especially those caused by displaced tracks from additional $pp$
interactions, \textit{i.e.}, pile-up events. 
Moreover, a pixel detector with a reduced pixel size of $25\times
100~\mu\text{m}^2$ is being developed \cite{Lange:2015acq}, which may improve the
momentum resolution for a disappearing track in future colliders.

Taking these potential improvements into account, in the following
analysis, we use a momentum cut of $P^{\text{dis}}_{\text{T}} > 100$~GeV
and assume that the number of BG events is not so much increased
due to deterioration in the momentum resolution.

%%%%%%%%%%%%%%%%%%
\section{Result}
%%%%%%%%%%%%%%%%%%

\begin{figure}
%\centering
{\includegraphics[width=0.45\textwidth]{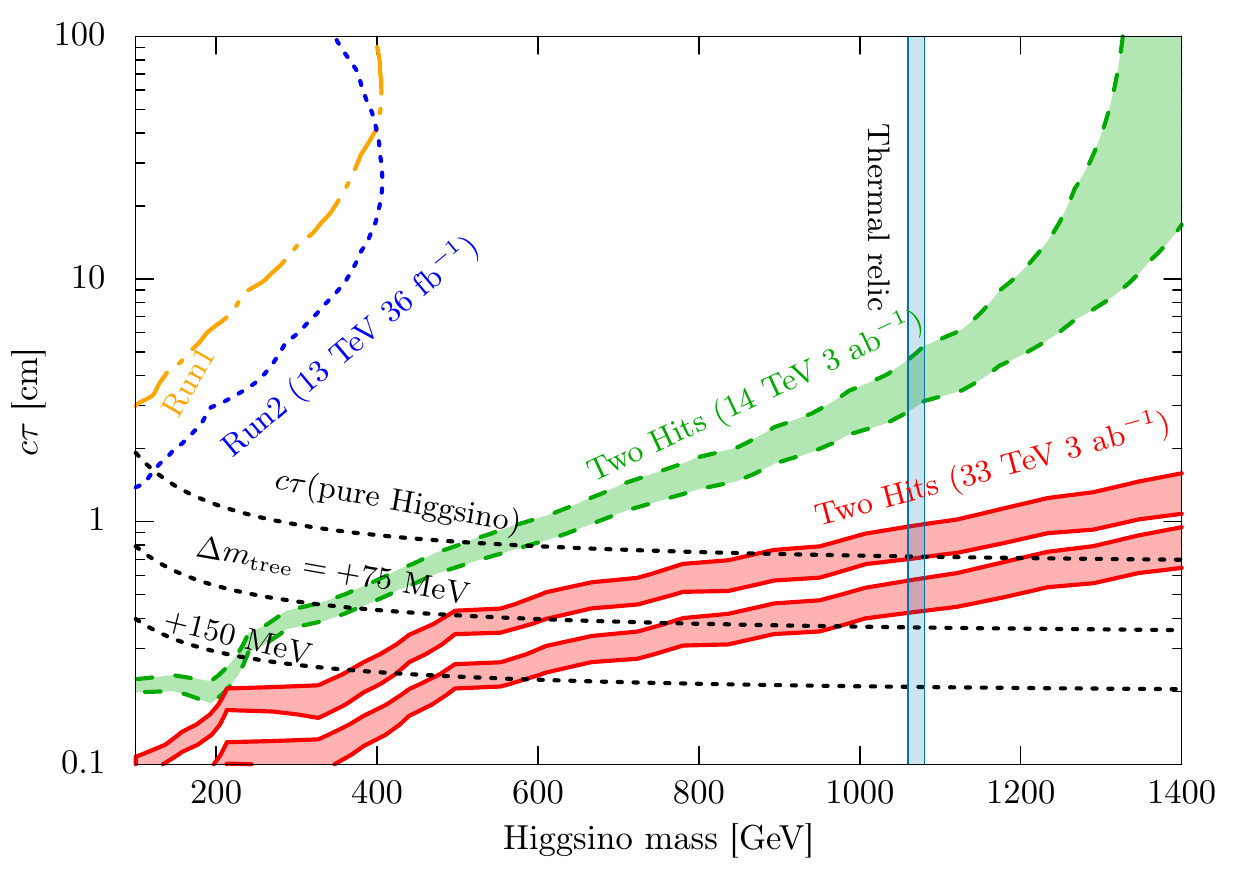}} 
\caption{
Expected limits on Higgsino from
 disappearing track searches at the LHC and a 33~TeV collider.
}
\label{fig:higgsino}
\end{figure}

Now we discuss the prospects of our two-hit method for the long-lived
charged Higgsino searches. 
We use {\sc Madgraph5} \cite{Alwall:2014hca}, {\sc Pythia6}
\cite{Sjostrand:2006za,*Sjostrand:2007gs}, and {\sc Delphes3}
\cite{deFavereau:2013fsa} for collider simulations and estimate the
production cross sections of SUSY particles with {\sc Prospino2}
\cite{Beenakker:1996ed}. 
We consider the 8~TeV LHC with
$20.3$~fb$^{-1}$ data \cite{Aad:2013yna}, the 13~TeV run with
$36.1$~fb$^{-1}$ data \cite{moriond17, *moriond17conf}, and the 14-TeV LHC and a future
33~TeV run with an integrated luminosity of $3$~ab$^{-1}$. We adopt the
same set of kinematic selection criteria as in the Run-1 analysis
\cite{Aad:2013yna}, except for the missing energy cut,
$E^{\text{miss}}_{\text{T}} > 140$~GeV, and the leading-jet transverse
momentum cut, $P^{\text{lead}}_{\text{T}} > 140$~GeV for the 8 and
13~TeV cases. For the 14 and 33 TeV cases, we require
$E_{\text{T}}^{\text{miss}}, P^{\text{lead}}_{\text{T}} > 400$~GeV and
600~GeV, respectively. As mentioned above, for the selection of disappearing
track candidates, we require two hits in the silicon tracker with
$P^{\text{dis}}_{\text{T}} > 100$~GeV. To estimate the number of
BG events for each case, we rescale the observed number of
BG events in the ATLAS Run-2 study \cite{moriond17, *moriond17conf} according to
the event rates of the $W+\text{jets}$ and $t\bar{t}$ processes. We find
that the expected number of BG events is $\sim 10$ for these
cases. 

In Fig.~\ref{fig:higgsino}, we show the expected limits from the 8~TeV,
13~TeV, 14~TeV, and 33~TeV searches in the yellow dot-dashed line, the blue
dashed line, the green dashed band, and the red bands, respectively. For the
33~TeV run, we assume that the second layer of a pixel detector is
located at a radius of 5 and 3~cm for the upper and lower bands,
respectively. The upper (lower) line of each band corresponds to ten
(zero) BG events. The systematic error in BG expectation is assumed to be $10~\%$. In addition, we show the decay lengths of Higgsino with
$\Delta m_{\text{tree}} = 0$, 75, and 150~MeV in black dotted lines
from top to bottom. The vertical blue stripe at $\sim 1.1$~TeV indicates the
favored mass value for pure Higgsino DM in terms of thermal
relic abundance \cite{Cirelli:2007xd}. This figure shows that the Run-2
data already have a sensitivity to pure Higgsino. The reach of the
high-luminosity LHC is expected to be 500--600 GeV. Moreover, a future
33~TeV $pp$ collider may probe pure Higgsino with a mass of $\gtrsim
1$~TeV, with which we can test the pure Higgsino DM scenario.

\begin{figure}
%\centering
{\includegraphics[width=0.45\textwidth]{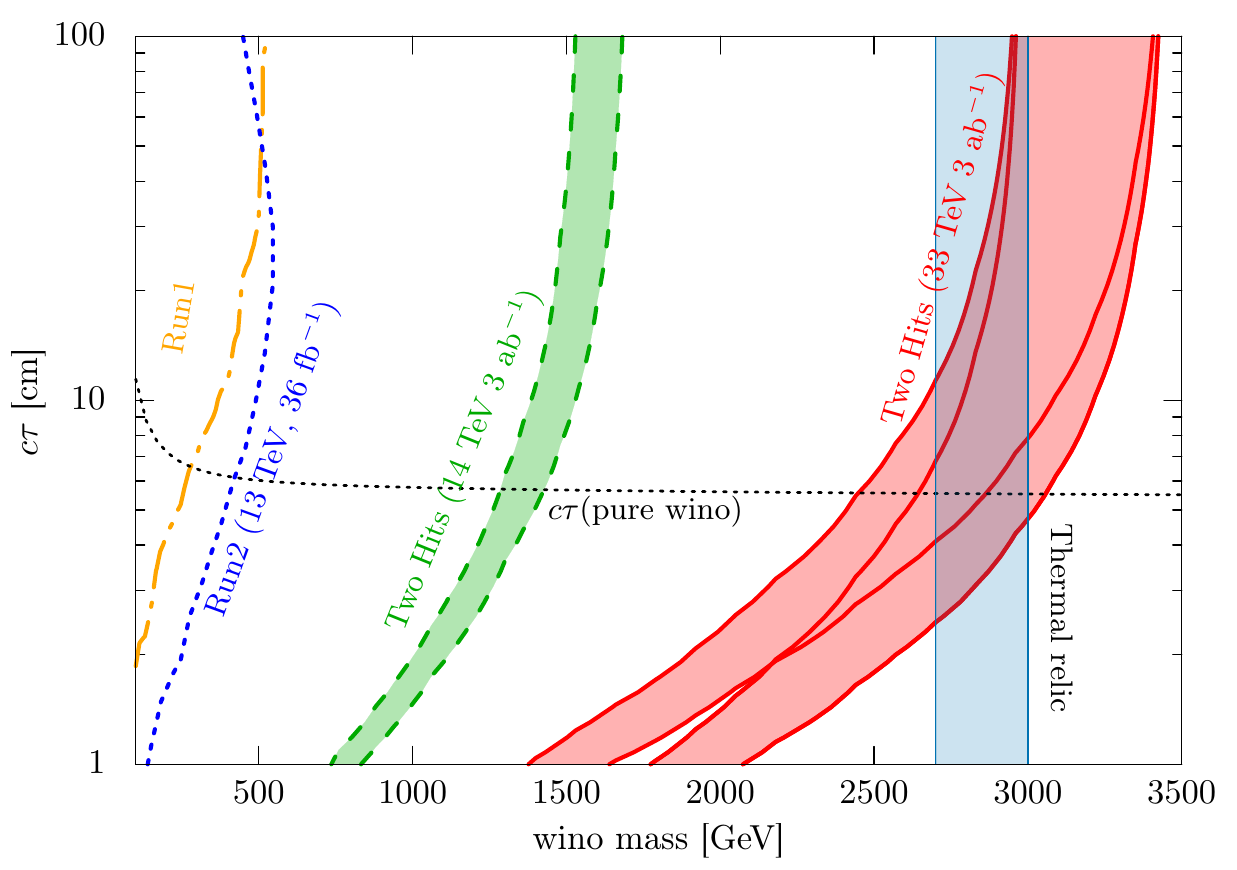}} 
\caption{
The current \cite{Aad:2013yna, moriond17, *moriond17conf} and expected limits on wino
 at the LHC and a 33~TeV collider. 
}
\label{fig:wino}
\end{figure}

Our two-hit search strategy can also extend the reach of charged wino
searches. To see this, we perform a similar analysis for the wino LSP
and show the results in Fig.~\ref{fig:wino}. The colors and types of the
lines and stripes are the same as in Fig.~\ref{fig:higgsino}.
The 8~TeV and 13~TeV limits in this figure are taken from
Refs.~\cite{Aad:2013yna, moriond17, *moriond17conf}. It is found that the reach of the
high-luminosity LHC for a pure wino can exceed 1~TeV. A future 33~TeV
collider may probe a pure wino with a mass of $\gtrsim 2.5$~TeV, where
(the most of) the observed DM density is explained by the
thermal relic abundance of the wino LSP \cite{Hisano:2006nn}. 

Finally, we note in passing that our two-hit search may probe Higgsino
and wino up to 1.7~TeV and 4~TeV or even larger, respectively, at a
100~TeV collider with an integrated luminosity of 3~ab$^{-1}$.

%%%%%%%%%%%%%%%%%%%%%%%%%%%%%%
\section{Conclusion and discussion}
%%%%%%%%%%%%%%%%%%%%%%%%%%%%%%

In this letter, we have discussed the testability of the Higgsino DM
scenario. We have seen that a wide region of parameter space for the
Higgsino DM scenario can be probed in future non-accelerator
experiments. However, if the electroweak-inos are so heavy that the
tree-level charged-neutral Higgsino mass difference is less than ${\cal
O}(100)$~MeV, Higgsinos may escape from these non-accelerator
searches. To probe such cases, in this paper, we propose a new collider
search strategy for almost pure Higgsinos based on the first two layers
of the Pixel detector, which is sensitive to ${\cal O}(1)$~cm disappearing
tracks. We have found that using this two-hit search method we can probe
1~TeV pure Higgsino, whose thermal relic abundance agrees with
$\Omega_{\text{DM}} h^2 \simeq 0.12$, at a 33~TeV collider. As a
consequence, the non-accelerator experiments and the disappearing track
searches play a complementary role in probing Higgsinos, and thus the
interplay among these experiments is of great importance to test the
Higgsino DM scenario experimentally. 

This method can also improve the disappearing track searches for
long-lived charged winos. In fact, since pure charged winos have
relatively long decay length, $\sim 6$~cm, we expect a number of signal
events with the two-hit search strategy. In this case, we may even
require two disappearing tracks, which can reduce the SM BG
significantly. We also note that this two-hit strategy is useful
for the search of other electroweak-charged DM candidates. A detailed study
of these searches, as well as the reduction of BG by means of DVs, will
be discussed elsewhere \cite{FNOS}. 

As we have seen, our proposal for the detection of disappearing tracks 
extends the LHC reach of the Higgsino and wino DM searches significantly.
Nonetheless, the mass values favored by thermal relic are beyond the 
LHC reach, and thus we need a new hadron collider to cover the entire region 
for the DM scenario. It is worth emphasizing that to maximize the potential 
of such a collider, not only an increase in the beam energy and luminosity,
but also an improvement in the tracker system is of crucial importance, 
which should be considered seriously when we discuss proposals for future colliders.

\vspace{2mm}
\noindent
{\it Note Added:} 
During the completion of this paper, a related study
\cite{Mahbubani:2017gjh} was submitted to arXiv where pure Higgsino
searches based on disappearing tracks at the LHC and a future 100~TeV
collider were discussed on the assumption that a charged track with a
length of 10~cm is detectable. We also noticed that the ATLAS
collaboration started to study a possibility of using a
two-hit strategy to search for long-lived charged winos \cite{jps17}.

\vspace{2mm}

%%%%%%%%%%%%%%%%%% Acknowledgements %%%%%%%%%%%%%%%%%%%%%%%%%%%%%%%%%%%%%%%
\begin{acknowledgments}
We are grateful to M. Saito for helpful correspondence. 
The work of H.O. was supported by JSPS KAKENHI Grant Number 15K17653.
The work of S.S. was supported by World Premier International Research Center
Initiative (WPI), MEXT, Japan.
The work of H.F. is supported
in part by a Research Fellowship for Young Scientists
from the Japan Society for the Promotion of Science
(JSPS).

\end{acknowledgments}

%%%%%%%%%%%%%%%%%%%%%%%%%%%%%%%%%%%%%%%%%%%%%%%%%%%%%%%%%%%%%%%%%%%%%%%%%%%

%%%%%%%%%%%%%%%%% Ref %%%%%%%%%%%%%%%%%%%%%%%%
\bibliography{ref}
%%%%%%%%%%%%%%%%%%%%%%%%%%%%%%%%%%%%%%%%%%%%%%

\end{document}